\documentclass[prl,showpacs,preprintnumbers,amsmath,aps,twocolumn]{revtex4}
\usepackage{amssymb}
\usepackage{amssymb}
\usepackage{amssymb}
\usepackage{txfonts}
\usepackage{amssymb}
\usepackage{amssymb}

\newcommand{\bei}{\begin{itemize}}
\newcommand{\eei}{\end{itemize}}
\newcommand{\bee}{\begin{enumerate}}
\newcommand{\eee}{\end{enumerate}}

\begin{document}

\title{Adiabatic Approximation Condition}
\author{Jian-da Wu$^1$}
\email{jdwu@mail.ustc.edu.cn}
\author{Mei-sheng Zhao$^1$, Jian-lan Chen$^1$ and Yong-de Zhang$^{2,1}$}

\affiliation{$^1$Hefei National Laboratory for Physical Sciences at
Microscale and Department of Modern Physics, University of Science
and Technology of China, Hefei 230026, People's Republic of China
\\$^2$CCAST (World Laboratory), P.O.Box 8730, Beijing 100080,
People's Republic of China}

\date{\today}

\begin{abstract}
In this paper, we present an invariant perturbation theory of the
adiabatic process based on the concepts of $U(1)$-invariant
adiabatic orbit and $U(1)$-invariant adiabatic expansion. As its
application, we propose and discuss new adiabatic approximation
conditions.

\end{abstract}


\maketitle
%---------------------------------------------------------------------------------------

Since the establishment of the quantum adiabatic theorem [1,2,3,4]
in 1923, many fundamental results have been obtained, such as
Landau-Zener transition [5], the Gell-Mann-Low theorem [6], Berry
phase [7] and holonomy [8]. Also the adiabatic processes find their
applications in the quantum control and quantum computation
[9,10,11,12]. Recently the common-used quantitative adiabatic
condition [15,16,17] has been found not able to guarantee the
validity of the adiabatic approximation [13,14]. Consequently
various new conditions are conjectured and a series of confusions
and debates arise. For example, it was argued [18] that the
traditional adiabatic condition did not have any problem at all and
that the invalidation of the condition did not mean the invalidation
of adiabatic theorem [19]. Some new conditions proposed in [20,21]
but too rigorous to be used conveniently. Although [22] also adopted
the adiabatic perturbation expansion but did not give out proper
condition because the basis in [22] can not show certain geometric
properties in the adiabatic process. [23] pointed out the limitation
of traditional condition but also did not give out a proper
condition. To solve the problem of insufficiency of traditional
adiabatic condition in [13,14] and clarify the subsequent
confusions, we introduce the concepts of $adiabatic$ $orbit$,
$U(1)$-$invariant$ $adiabatic$ $orbit$ and $U(1)$-$invariant$
$adiabatic$ $evolution$ $orbit$. The meanings of adiabatic evolution
is reclaimed. And new adiabatic approximation conditions based on
the $U(1)$-$invariant$ $adiabatic$ $expansion$ with the
time-dependent coefficient are proposed and illustrated by two
examples.

Let us consider a quantum system governed by a time dependent
Hamiltonian $H(t)$ and the initial state of the system is an
eigenstate $|m,0\rangle$ of $H(0)$ with eigenvalue $E_m (0)$, where
$m$ denotes the initial value of dimensionless quantum number set.
By introducing a dimensionless time parameter $\tau = E_m \left( 0
\right)t/\hbar $ and a dimensionless Hamiltonian $h(\tau ) = H(\tau
)/E_m \left( 0 \right)$,  the time dependent Schr\"odinger equation
reads

\begin{equation}\label{g1}
 i\frac{\partial |\Phi_m(\tau)\rangle}{\partial \tau }
 = h(\tau )|\Phi_m(\tau)\rangle,\quad
 \left|\Phi_m( 0)\right\rangle = \left| {m,0} \right\rangle.
\end{equation}
The exact solution $\left| {\Phi_m \left( \tau \right)}
\right\rangle$ to Eq.(\ref{g1}) is referred to as the system's
$dynamic$ $evolution$ $orbit$ in the Hilbert space.

Furthermore, by considering $\tau$ as a fixed parameter, we can
always solve the following quasi-stationary equation of the
Hamiltonian $h\left( \tau  \right)$

\begin{equation}
h\left( \tau  \right)\left| {\varphi _n \left( \tau  \right)}
\right\rangle  = e_n \left( \tau  \right)\left| {\varphi _n \left(
\tau  \right)} \right\rangle. \label{g2}
\end{equation}
And the eigenstate $|\varphi_n(\tau)\rangle$ with the corresponding
initial state $\left| {n,0} \right\rangle$ is referred to as the
$adiabatic$ $solution$ or the $adiabatic$ $orbit$ of the system.

For convenience, we denote $\gamma _{nm}  \equiv i\left\langle
{{\varphi _n (\tau )}}
 \mathrel{\left | {\vphantom {{\varphi _n (\tau )} {\dot \varphi _m (\tau )}}}
 \right. \kern-\nulldelimiterspace}
 {{\dot \varphi _m (\tau )}} \right\rangle$ and the dot here and below
expresses the derivative with respect to time. Apparently, an
adiabatic orbit multiplied by an arbitrary time-dependent phase
factor still describes the same adiabatic orbit. It is not difficult
to see that the following adiabatic orbit
\begin{equation}
\left| {\Phi _m^{adia} (\tau)} \right\rangle  = \exp\left\{ -
i\int_0^\tau  [e_m (\lambda ) -\gamma _{mm}(\lambda)]d\lambda
\right\}| {\varphi _m (\tau )}\rangle \label{a4}
\end{equation}
is invariant [26] under the following $U(1)$ transformation
\begin{equation}\label{gauge}
\left| {\varphi _m (\tau )} \right\rangle  \to e^{if_m (\tau )}
\left| {\varphi _m (\tau )} \right\rangle \;\;(f_m (0) = 0).
\end{equation}
Here $f_m (0) = 0$ is because of given initial state. We call this
adiabatic orbit with special choice of the time-dependent phase
factor as the $U(1)$-$invariant$ $adiabatic$ $orbit$.

It is clear that, although the initial conditions $\left| {m,0}
\right\rangle$ are the same, the dynamic evolution orbit
$|\Phi_m(\tau)\rangle$ do not always coincide with the adiabatic
orbit $|\varphi_m(\tau)\rangle$, or they are not even close to each
other. Obviously they coincide if and only if

\begin{equation}
\gamma _{nm} = 0\;\;(\forall n \ne m). \label{a3}
\end{equation}
In this case, Eq.(\ref{a4}) being the solution to both Eq.(\ref{g1})
and Eq.(\ref{g2}) is referred to as the $U(1)$-$invariant$
$adiabatic$ $evolution$ $orbit$, describing a strict adiabatic
evolution orbit of the system.

Generally speaking, the dynamic evolution orbit $\left| {\Phi _m
\left( \tau  \right)} \right\rangle$ starting from the initial state
$\left| {m,0} \right\rangle $ will change among some adiabatic
orbits which will cause transitions between different them. Our task
is to find the proper condition under which the dynamic orbit is
sufficiently close to the adiabatic orbit when the Eq.(\ref{a3}) is
not satisfied.

Since the Hamiltonian $h(\tau)$ is Hermitian, all the
$U(1)$-$invariant$ $adiabatic$ $orbits$ in Eq.(\ref{a4}) at a given
time constitute a complete orthonormal basis of the system. In this
basis, the dynamic evolution orbit of system reads
\begin{equation}
\left| {\Phi _m \left( \tau  \right)} \right\rangle  = \sum\limits_n
{c_n (\tau )\left| {\Phi _n^{adia} \left( \tau  \right)}
\right\rangle \;,\;\;\;\;}  {\left| {\Phi _m \left( 0 \right)}
\right\rangle }  = \left| {m,0} \right\rangle. \label{a6}
\end{equation}
The expansion in Eq.(\ref{a6}) is referred to as the
$U(1)$-$invariant$ $adiabatic$ $expansion$ with the time-dependent
coefficients. Therefore, the set of coefficients equations reads

\begin{equation}
\dot c_m (\tau ) = i\sum\limits_{n \ne m} {c_n (\tau )M(\tau )_{mn}
} \;,\label{t7}
\end{equation}
where the diagonal elements of the marix $M(\tau )$ are zero and the
non-diagonal elements of $M(\tau )$ read
\begin{equation}
M(\tau )_{mn}  = i\left\langle {{\Phi _m^{adi} (\tau )}}
 \mathrel{\left | {\vphantom {{\Phi _m^{adi} (\tau )} {\dot \Phi _n^{adi} (\tau )}}}
 \right. \kern-\nulldelimiterspace}
 {{\dot \Phi _n^{adi} (\tau )}} \right\rangle  \equiv \left| {\gamma _{mn} (\tau )} \right|e^{i\theta _{mn} (\tau
 )}, \label{t8}
 \end{equation}
where
\begin{equation}
\theta _{mn} (\tau ) = \int_0^\tau  {d\lambda \left( {e_m (\lambda )
- e_n (\lambda ) + \gamma _{nn}  - \gamma _{mm} } \right)}  + \arg
\gamma _{mn} (\tau ).
 \end{equation}

Thus, the probability of staying in adiabatic orbit $\left| {\Phi
_m^{adia} \left( \tau  \right)} \right\rangle$ is

\begin{equation}
P_m \left( \tau  \right) = \left| {c_m \left( \tau  \right)}
\right|^2  = \left| {\left( {\hat T\exp \left[ {i\int_0^\tau
{d\lambda M\left( \lambda  \right)} } \right]} \right)_{mm} }
\right|^2, \label{g7}
 \end{equation}
where $\hat T$ is time ordered operator. And one can obtain further
detailed analysis on Eq.(\ref{g7}) in our another paper [24].

Accordingly, the adiabatic approximation of system requires

\begin{equation}
P_m \left( \tau  \right) \to 1. \label{g8}
\end{equation}
It means the transition probability from dynamic evolution orbit to
other adiabatic orbits (except the adiabatic orbit $\left| {\Phi
_m^{adia} \left( \tau  \right)} \right\rangle$) can be neglected.

According to the perturbation theory for Eq.(\ref{g7}), the
first-order approximation of $P_m(\tau)$ is
\begin{equation}
P_m (\tau ) = 1 - \sum\limits_{n \ne m} {\left| {\int_0^\tau {\left|
{\gamma _{nm} (\lambda )} \right| e^{i\theta _{nm} (\lambda )}
d\lambda } } \right|^2 }. \label{a9}
 \end{equation}
Therefore, the adiabatic approximation requires

\begin{equation}
\left| {\int_0^\tau  {\left| {\gamma _{nm} (\lambda ) } \right|
e^{i\theta _{nm} (\lambda )} d\lambda } } \right|^2  \to
0\;\;(\forall n \ne m). \label{g9}
 \end{equation}
For general situation $\left| {\theta_{nm} (\tau ) - \theta_{nm}
(0)} \right| \ge 2\pi$ and $\left| {\dot \theta_{nm} (\tau )}
\right| \ge 1$, the integral of Eq.(\ref{g9}) will be sufficiently
small,  if the phase of the integrated function vibrates fast enough
and the amplitude  of the integrated function is small enough, thus
we will have following adiabatic condition

\begin{equation}
\left| {\dot \theta _{nm} (\tau )} \right| \gg \left| {\gamma _{nm}
(\tau )} \right| \;\;\;\;(\forall n \ne m)\label{m11}
 \end{equation}
that is
\begin{equation}
\left| {e_n (\tau ) - e_m (\tau ) + \Delta _{mn} (\tau )} \right|
\gg \left| {\gamma _{nm} (\tau )} \right|\;\;\;(\forall n \ne m)
\label{g11}
\end{equation}
where
\begin{equation}
\Delta _{mn} \left( \tau  \right) \equiv  {\gamma _{mm} \left( \tau
\right) - \gamma _{nn} \left( \tau  \right) + \frac{d}{{d\tau }}\arg
\gamma _{nm} \left( \tau  \right)}\;\;(\forall n \ne m). \label{c13}
\end{equation}
Here $\Delta _{mn}$ referred to as $quantum$ $geometric$ $potential$
is a new quantity comparing to the traditional adiabatic condition.
And it should be noticed [26] that the $quantum$ $geometric$
$potential$ is also $U(1)$-invariant under the transformation
Eq.(\ref{gauge}). It should point out that $\Delta _{mn}$ appears
naturally in our theory. And for $quantum$ $geometric$ $potential$
one can obtain further detailed analysis and application in our
another paper [25]. Furthermore, from Eq.(\ref{g9}), the change of
the phase in the integrated function should be much larger than the
integral of amplitude, we will have another condition in integral
form

\begin{equation}
\left| {\int_0^\tau  {d\lambda } \left[ {e_n (\lambda ) - e_m
(\lambda ) + \Delta _{mn} } \right]} \right| \gg \int_0^\tau
{d\lambda \left| {\gamma _{nm} \left( \lambda  \right)} \right|}
,\forall \;n \ne m. \label{g12}
\end{equation}
Based on Eq.(\ref{t7}) and Eq.(\ref{t8}), we also prove [27] a
theorem related to condition Eq.(\ref{m11}) and Eq.(\ref{g12}).
Following analysis and our subsequent works [24,25] indicate that
Eq.(9-18) are of abundant content.

Next we will give two examples to show the validity of
Eq.(\ref{g11}) and Eq.(\ref{g12}).

The first example is to indicate that the problems shown in [13,14]
do not exist because the relation between system $a$ and $b$ stated
in Ref.[13,14] does not guarantee our condition Eq.(\ref{g11}). The
proof is given below.

[14] showed that for an arbitrary time-dependent system $a$ with
Hamiltonian $h^a (\tau )$ and quasi-stationary equation $h^a (\tau
)\left| {n^a (\tau )} \right\rangle  = e_n^a (\tau )\left| {n^a
(\tau )} \right\rangle$, one can construct time-dependent system $b$
with Hamiltonian $h^b (\tau )$ and quasi-stationary equation $h^b
(\tau )\left| {n^b (\tau )} \right\rangle  = e_n^b (\tau )\left|
{n^b (\tau )} \right\rangle$ as follows

\begin{equation}
\left\{ {\begin{array}{*{20}c}
   {h^a (\tau ) = i\dot U(\tau )U^\dag  (\tau ),h^b (\tau ) = i\dot U^\dag  (\tau )U(\tau )}  \\
   {\;\left| {n^b (\tau )} \right\rangle  = U^\dag  (\tau )\left| {n^a (\tau )} \right\rangle ,\quad e_n^b (\tau ) =  - e_n^a (\tau )}.  \\
\end{array}} \right. \label{c22}
\end{equation}
Simple calculation yields
\begin{equation}
\begin{array}{l}
 \gamma _{nm}^b \left( \tau  \right) = -  e_m^a (\tau )\delta _{mn}  + \gamma _{nm}^a \left( \tau  \right) \\
 \end{array}. \label{g13}
\end{equation}
Using Eq.(\ref{g11}), we have

\begin{equation}
\frac{{\left| {\gamma _{nm}^a \left( \tau  \right)}
\right|}}{{\left| {e_m^a (\tau ) - e_n^a (\tau ) - \Delta _{mn}^a
\left( \tau  \right)} \right|}} \ll 1,\quad \forall n \ne m
\label{g14}
\end{equation}
for system $a$, while for system $b$, we will have

\begin{equation}
\frac{{\left| {\gamma _{nm}^b \left( \tau  \right)}
\right|}}{{\left| {e_m^b (\tau ) - e_n^b (\tau ) - \Delta _{mn}^b
\left( \tau  \right)} \right|}} = \frac{{\left| {\gamma _{nm}^a
\left( \tau  \right)} \right|}}{{\left| {\Delta _{mn}^a \left( \tau
\right)} \right|}}. \label{g15}
 \end{equation}
Comparing adiabatic conditions between system $a$ Eq.(\ref{g14}) and
system $b$ Eq.(\ref{g15}), the denominator of Eq.(\ref{g14}) has an
extra term $e_m^a (\tau ) - e_n^a (\tau )$. Therefore, we can not
conclude that system $b$ still satisfies Eq.(\ref{g11}). This
indicates that the problems stated in [13,14] will not exist if we
adopt the new adiabatic condition Eq.(\ref{g11}). And it is
worthwhile to point out that we can present accurate adiabatic
condition for the examples given in [14] according to our condition
Eq.(\ref{g11}).

The second example is to consider the well-known model, a spin-half
particle in a general magnetic field. The Hamiltonian of the system
is

\begin{equation}
 h(\tau ) = \eta \sigma _z  + \xi (\tau )\left( {\sigma _x \cos 2\eta
\tau  + \sigma _y \sin 2\eta \tau } \right) \label{g16}
\end{equation}
where $\eta  = \hbar {{\omega _0 } \mathord{\left/
 {\vphantom {{\omega _0 } {E_ \pm  (0)}}} \right.
 \kern-\nulldelimiterspace} {E_ \pm  (0)}}$, ${{\xi (\tau ) = \hbar \omega \left( \tau  \right)} \mathord{\left/
 {\vphantom {{\xi (\tau ) = \hbar \omega \left( \tau  \right)} {E_ \pm  (0)}}} \right.
 \kern-\nulldelimiterspace} {E_ \pm  (0)}}$ and $\omega _0$ is
 a constant. Suppose that the initial state of the system is $\left| { \pm
,0} \right\rangle = \exp \left( { - i\sigma _y \theta (0)/2}
\right)\left| {
\pm\mathord{\buildrel{\lower3pt\hbox{$\scriptscriptstyle\rightharpoonup$}}
\over e} _z } \right\rangle$ with energy eigenvalues $E_ \pm (0)$ at
initial time respectively, where $\theta (0) = \arctan (\omega
(0)/\omega _0 )$ and $\left| {
\pm\mathord{\buildrel{\lower3pt\hbox{$\scriptscriptstyle\rightharpoonup$}}
\over e} _z } \right\rangle$ are the eigenstates of $\sigma _z$. The
dynamic evolution orbits of the system are

\begin{equation}
\left| {\Phi _ \pm  (\tau )} \right\rangle  = e^{ - i\sigma _z \eta
\tau } e^{ - i\sigma _x \int_0^\tau  {\xi (\lambda )d\lambda } }
\left| { \pm ,0} \right\rangle. \label{g17}
 \end{equation}
Here, the two $U(1)$-invariant adiabatic orbits passing through the
corresponding initial eigenstates $\left| { \pm ,0} \right\rangle$
are

\begin{equation}
\left| {\Phi _ \pm ^{adia} \left( \tau  \right)} \right\rangle  =
e^{ - i\eta \tau \sigma _z } e^{ - i\theta (\tau )\sigma _y /2} e^{
- i\sigma _z \int_0^\tau  {\left( {  \Omega (\lambda ) - \frac{{\eta
^2 }}{{\Omega (\lambda )}}} \right)d\lambda } } \left| { \pm
\mathord{\buildrel{\lower3pt\hbox{$\scriptscriptstyle\rightharpoonup$}}
\over e} _z } \right\rangle, \label{g18}
\end{equation}
here $\Omega (\tau ) = \sqrt {\xi (\tau )^2  + \eta ^2 }$, $\cos
\theta (\tau ) = \eta /\Omega (\tau ) $
 and $\left| {\Phi _ \pm ^{adia} (0)} \right\rangle  = \left| { \pm ,0} \right\rangle$.

Now, we use the new adiabatic condition Eq.(\ref{g12}) to examine
under what circumstances the evolution of the system keeps in the
adiabatic orbit, $\left| {\Phi _ + ^{adia} \left( \tau \right)}
\right\rangle$. First of all, it is easy to calculate the
probability of finding the dynamic evolution orbit of the system in
the adiabatic orbit $\left| {\Phi _ + ^{adia} \left( \tau  \right)}
\right\rangle$

\begin{equation}
P_m = \left| {\left\langle {{\Phi _ + ^{adia} \left( \tau  \right)}}
 \mathrel{\left | {\vphantom {{\Phi _ + ^{adia} \left( \tau  \right)} {\Phi _ +  (\tau )}}}
 \right. \kern-\nulldelimiterspace}
 {{\Phi _ +  (\tau )}} \right\rangle } \right|^2  = \frac{1}{2} +
 \frac{1}{2}\frac{{\xi (0)\xi (\tau ) + \eta ^2 \cos 2\delta }}{{\Omega (0)\Omega (\tau )}}, \label{g20}
\end{equation}
where $\delta  =  - \int_0^\tau  {\xi (\lambda )d\lambda }$.

Suppose that $\left| {\dot \theta _{ -  + } (\tau )} \right| \ge 1$
and $\left| {\theta _{ -  + } (\tau ) - \theta _{ -  + } (0)}
\right| \ge 2\pi$, then
 according to the condition Eq.(\ref{g12}), we will have

\begin{equation}
\begin{array}{l}
 \left| {2\int_0^\tau  {d\lambda \frac{{\xi (\lambda )^2 }}{{\Omega (\lambda )}}}  - \left( {\left. {\arg \gamma _{ -  + } (\lambda )} \right|_0^\tau  } \right)} \right| \\
  \gg \int_0^\tau  {d\lambda \frac{{\eta \xi (\lambda )}}{{\Omega (\lambda )}}\sqrt {\frac{{\dot \xi (\lambda )^2 }}{{4\Omega (\lambda )^2 \xi (\lambda )^2 }} + 1} }.  \\
 \end{array}
\label{g22}
\end{equation}
After simple analysis, we will obtain following sufficient condition

\begin{equation}
\xi (\tau ) \gg \eta  \label{g23}
\end{equation}
or

\begin{equation}
\eta  \gg \xi (\tau )\;\;and\;\int_0^\tau  {\xi (\lambda )d\lambda }
\ll 1. \label{g24}
\end{equation}
It is easy to see that when Eq.(\ref{g23}) or Eq.(\ref{g24}) is
satisfied, from Eq.(\ref{g20}), we have

\begin{equation}
P_m = \left| {\left\langle {{\Phi _ + ^{adia} \left( \tau
\right)}}
 \mathrel{\left | {\vphantom {{\Phi _ + ^{adia} \left( \tau  \right)} {\Phi _ +  (\tau )}}}
 \right. \kern-\nulldelimiterspace}
 {{\Phi _ +  (\tau )}} \right\rangle } \right|^2  \approx 1. \label{g25}
\end{equation}
Namely, our new adiabatic condition Eq.(\ref{g12}) guarantees the
evolution of the system is an adiabatic evolution. What is more, we
can choose proper $\xi (\tau )$ to be periodic, so the condition
stated in Ref.[20] can not be applied to our example and has obvious
limitation. In fact, the adiabatic process may be a longtime
vibration process, so the condition influenced by the times of
vibration stated in Ref.[20] is too much strict. As for the general
sufficient condition stated in Ref.[21], it is not only too
complicated to operate but also too rigorous to apply.

In conclusion, according to the concepts of adiabatic orbit,
$U(1)$-invariant adiabatic orbit and adiabatic evolution orbit
stated in our paper, we reclaim the meanings of adiabatic evolution
and present an invariant perturbation theory of adiabatic process
based on time-dependent $U(1)$ invariant adiabatic expansion.  Of
course, Eq.(\ref{g11}) and Eq.(\ref{g12}) can not be proved to be
sufficient condition, thus we give out sufficient conditions in [27]
which contains Eq.(\ref{g11}) and Eq.(\ref{g12}), however, the
second condition in [27] are too strict to exclude many interesting
physical systems. As far as we know, the conditions Eq.(\ref{g11})
and Eq.(\ref{g12}) can be used to determine whether the evolution of
the system is adiabatic or not for all familiar examples listed in
the existed papers.  We also preliminarily show the influence of
quantum geometric potential in the new adiabatic condition. Further
detailed discussions on quantum geometric potential and the new
adiabatic conditions can be seen in [24,25].

\acknowledgements

We thank Prof. Sixia Yu and Dr. Dong Yang for illuminating
discussions. This work is supported by the NNSF of China, the CAS,
and the National Fundamental Research Program (under Grant No.
2006CB921900).

%---------------------------------------------------------------------------------------

\begin{eqnarray}
c_m(T)-1=i\sum_{n\ne m}\int_0^Td\tau
\;|\gamma_{mn}(\tau)|e^{-i\theta_{mn}(\tau)}c_n(\tau)\cr =\sum_{n\ne
m}\frac{{\left| {\gamma _{mn} } \right|}}{{\dot \theta _{mn}
}}e^{i\theta_{mn}}c_n(\tau)\Big|_0^T
    -\sum_{n\ne
m}\int_0^Td\tau\frac{d}{{d\tau }}\left( {\frac{{\left| {\gamma _{mn}
} \right|}}{{\dot \theta _{mn} }}} \right)
e^{i\theta_{mn}}c_n(\tau)\cr -i\sum_{n\ne m,l\ne
n}\int_0^Td\theta_{nl}(\tau)\frac{{\left| {\gamma _{mn} }
\right|}}{{\dot \theta _{mn} }}\frac{{\left| {\gamma _{nl} }
\right|}}{{\dot \theta _{nl} }}e^{i(\theta_{mn}+\theta_{nl}(\tau))}
c_l(\tau).
\end{eqnarray}
Then from Eq.(33), Eq.(34) and Eq.(35), we have
\begin{equation}
1 - \left| {c_m (\tau )} \right| \le \left| {1 - c_m (\tau )}
\right| \le \varepsilon
\end{equation}
namely,
\begin{equation}
P_m  = \left| {c_m (\tau )} \right|^2  \ge \left( {1 - \varepsilon }
\right)^2.
\end{equation}
Thus we prove the theorem.


\begin{thebibliography}{99}
\bibitem{U(1)}
Here we will prove that $adiabatic$ $basis$ $\left\{ {\left| {\Phi
_m^{adia} \left( \tau  \right)} \right\rangle } \right\}$ and
$quantum$ $geometric$ $potential$ $\Delta _{mn} \left( \tau \right)$
are both invariant under arbitrary time-dependent $U(1)$
transformation.

$Proof$: Perform an arbitrary time-dependent $U(1)$ transformation
on adiabatic orbit $\left| {\varphi _n \left( \tau  \right)}
\right\rangle$
\begin{equation}
\left| {\varphi _n (\tau )} \right\rangle  \to \left| {\varphi '_n
(\tau )} \right\rangle  = e^{i\,f_n (\tau )} \left| {\varphi _n
(\tau )} \right\rangle.
\end{equation}
Thus, we will have

\begin{eqnarray}
\begin{array}{l}
 e^{ - i\int_0^\tau  {\left[ {e_m (\lambda ) - \gamma '_{mm} (\lambda )} \right]d\lambda } } \left| {\varphi '_m (\tau )} \right\rangle  \\
  = e^{if_m (0)} e^{ - i\int_0^\tau  {\left[ {e_m (\lambda ) - \gamma _{mm} (\lambda )} \right]d\lambda } } \left| {\varphi _m (\tau )} \right\rangle  \\
 \end{array}
 \end{eqnarray}
The given phase of initial state $\left| {m,0} \right\rangle $
requires $f_m \left( 0 \right) = 0$. Thus we prove adiabatic basis
$\left\{ {\left| {\Phi _m^{adia} \left( \tau  \right)} \right\rangle
} \right\}$ is $U(1)$-invariant. And the proof for $\Delta _{mn}
\left( \tau  \right)$ is similar. Performing the same $U(1)$
transformation, we will have

\begin{equation}
\begin{array}{l}
 \Delta '_{mn} (\tau ) = \left[ {\gamma _{mm} (\tau ) - \dot f_m (\tau )} \right] - \left[ {\gamma _{nn} (\tau ) - \dot f_n (\tau )} \right] \\
 \;\;\;\;\;\;\;\;\;\;\;\;\; + \frac{d}{{d\tau }}\arg \left[ {i\left\langle {\varphi _n (\tau )} \right|e^{ - if_n (\tau ) + if_m (\tau )} \left| {\left\langle {\varphi _m (\tau )} \right|} \right\rangle } \right] \\
 \;\;\;\;\;\;\;\;\;\; = \Delta _{mn} (\tau )\;\;\;\;(m \ne n) \\
 \end{array}
\end{equation}
These two proofs show that even if the phase $\gamma _{kk} \left(
\tau  \right)$ is dependent of time, we still can use them to
construct meaningful quantity which has physical meanings and will
not change under any time-dependent $U(1)$ transformation. The
quantum geometric potential $\Delta _{mn} \left( \tau  \right)$ and
adiabatic basis $\left\{ {\left| {\Phi _m^{adia} \left( \tau
\right)} \right\rangle } \right\}$ are just the cases. And all these
quantities should exist in all time-dependent process. We should
note that the traditional Berry phase can be included if we perform
cyclic integral on the phase in $\left\{ {\left| {\Phi _m^{adia}
\left( \tau \right)} \right\rangle } \right\}$.
\end{thebibliography}
\end{document}